\documentstyle[aps,epsfig,prl,amssymb,multicol,citesort]{revtex}

\def\unitR{{\mbox{\boldmath$\hat R$}}}

\def\f{{\mbox{\boldmath$f$}}}

\def\k{{\mbox{\boldmath$k$}}}

\def\v{{\mbox{\boldmath$v$}}}
\def\u{{\mbox{\boldmath$u$}}}
\def\U{{\mbox{\boldmath$U$}}}

\newcommand{\x}{{\mbox{\boldmath$x$}}}
\newcommand{\z}{{\mbox{\boldmath$z$}}}
\newcommand{\BR}{{\mbox{\boldmath$R$}}}

\newcommand{\SO}{{$SO(3)$}}
\newcommand{\be}{\begin{equation}}
\newcommand{\ee}{\end{equation}}
\newcommand{\bea}{\begin{eqnarray}}
\newcommand{\eea}{\end{eqnarray}}

\newcommand{\lp}{\left(}
\newcommand{\rp}{\right)}
\newcommand{\la}{\left\langle}
\newcommand{\ra}{\right\rangle}
\begin{document}
\title{Anomalous and dimensional scaling in anisotropic turbulence}
 \author{L.~Biferale$^{1,4}$, I.~Daumont$^{1}$, A.~Lanotte$^{2,4}$, and  F.~Toschi$^{3,4}$}
\address{$^1$ Dipartimento di Fisica, Universit\`a "Tor Vergata", Via della Ricerca Scientifica 1, I-00133 Roma, Italy}
\address{$^2$ CNR, ISAC - Sezione di Lecce, Str. Prov. Lecce-Monteroni Km 1.200, I-73100 Lecce, Italy}
\address{$^3$ CNR, Istituto per le Applicazioni del Calcolo, Viale del Policlinico 137, I-00161 Roma, Italy}
\address{$^4$ INFM, Unit\`a di Tor Vergata, Via della Ricerca Scientifica 1, I-00133 Roma, Italy}
\maketitle
\begin{abstract}
We present a numerical study of {\it anisotropic} statistical fluctuations in 
{\it  homogeneous} turbulent flows. We give an argument to predict the dimensional
scaling exponents, $\zeta^j_d(p) = (p+j)/3$, for the projections
of $p$-{\it th} order structure function in the $j$-{\it th} sector
of the rotational group. 
We show that measured exponents are anomalous, showing a clear 
deviation from the dimensional prediction. Dimensional scaling is 
subleading and it is recovered 
only after a random reshuffling of all velocity phases, 
in the stationary ensemble. This supports the idea that anomalous scaling is the 
result of a genuine inertial evolution, independent of large-scale 
behavior. 
\end{abstract}
\begin{multicols}{2}
In recent years a huge amount of theoretical,
 numerical and experimental work has been 
done in order to study anisotropic turbulent fluctuations 
\cite{ps95,p96,gw98,alp00,arad_exp,sk00,bt00,bv00,blmt02,gvl01,abmp99,sw01}.
Typical questions go from the theoretical point of calculating and measuring
anomalous scaling exponents in anisotropic sectors \cite{alp00,arad_exp,sk00,abmp99,bt00}, to the more applied
problem of quantifying the rate of recovery of isotropy 
at scales small enough \cite{bv00,gw98,p96,ps95}.
Another important issue is 
the universality of anisotropic scaling exponents, i.e. whether they are
an intrinsic characteristic of the Navier-Stokes non-linear evolution or
they are fixed by a dimensional matching with the external
anisotropic forcing.  

Important steps forward in the analysis of anisotropic fluctuations have 
recently been done in Kraichnan models, i.e. passive scalars/vectors advected 
by isotropic, Gaussian and white-in-time velocity fields 
\cite{rev}, with a large-scale anisotropic forcing 
\cite{lm99,abp00,arad_pass}. 
In those  models, anomalous scaling arises as the result of a non-trivial null-space 
structure for the advecting operator (zero modes). Also, correlation 
functions in different sectors of
the rotational group show different scaling properties. Scaling
exponents are universal: they do not depend on 
the actual value of the forcing and boundary conditions, and they 
are fully characterized by the order of the anisotropy. Non-universal
effects are felt only in coefficients multiplying the 
power laws.
 
Similar problems, like the existence of scaling laws in anisotropic
sectors and, if any, the values of the corresponding scaling exponents are at 
the forefront of experimental, numerical and theoretical research for real 
turbulent flows. 
Only few indirect experimental investigations of scaling in different
sectors \cite{arad_exp,sk00} and direct decomposition in numerical 
simulations \cite{abmp99,blmt02,bv00,bt00} have been attempted up to now. 

The question is still open, evidences
of a clear improving of scaling laws by isolating the isotropic sector
have been reported, supporting the idea that the undecomposed 
correlations are strongly affected by the superposition of isotropic
and anisotropic fluctuations \cite{abmp99,bt00}.
On a theoretical ground, only recently it has been highlighted
the potentiality of \SO\ decomposition 
to quantify different degrees of anisotropies for any correlation 
function \cite{alp00}. 
On the basis of this analysis, preliminary experimental evidences of the 
existence of a scaling law also in sectors with total angular
momentum $j=2$ have been reported \cite{arad_exp,sk00}. The value
of the exponent for the second order correlation function being close to the 
dimensional estimate $\zeta_d^{j=2}(2) = 4/3$,
\cite{lu67} (where, from now on, subscript $d$ denotes the dimensional value).

Tipically, experimental investigations in real
turbulent flows are flawed by the contemporary presence 
of anisotropies and strong non-homogeneities. 
The meaning of \SO\ decomposition becomes opaque in presence
of strong non-homogeneities and also the very existence of 
scaling laws cannot be given for granted \cite{fedeprl}.\\
To overcome such difficulties, some of us performed (see \cite{bt00}) 
the numerical investigation of a ``Random-Kolmogorov Flow'' (RKF), 
a fully periodic Kolmogorov flow with random forcing phases, 
$\delta$-correlated in time.

In this Letter, we present a more extended analysis of the same data set,
but focusing on new evidences that anisotropic scaling exponents are
indeed universal and anomalous, i.e. they do not follow simple dimensional 
scaling. 
In order to do this, we also give  a clear phenomenological background
able to predict the dimensional scaling in anisotropic sectors. 
The Letter is organized as follows. First, we present a simple dimensional
argument for all anisotropic sectors of structure functions of any order.
With respect to this dimensional prediction, we show that anisotropic 
exponents are indeed anomalous. 
Moreover, we show that by performing a random re-shuffling of the velocity 
phases (in the stationary ensemble of anisotropic configurations), the leading anomalous scaling is filtered out and the sub-leading 
dimensional prediction is recovered.
This is both a test of our dimensional prediction and 
a clean indication that forced velocity correlations are dominated 
by inertial terms; in this sense one may refer to them 
as the ``equivalent'' of the zero-modes responsible 
for anomalous scaling and universality in linear hydrodynamical problems 
\cite{rev}.
These findings leads to conclude that anisotropic fluctuations 
in turbulence are anomalous and universal.
\\
We recall few details on the numerical simulations \cite{bt00}.
The RKF is fully periodic; the large-scale 
anisotropic random forcing points in one direction, $\hat{\z}$,  
has a spatial dependency only from the $x$ coordinate and it is different from zero at the two wavenumbers: $\k_1=(1,0,0), \k_2=(2,0,0)$.  
Namely, $\f_i(\k_{\{1,2\}}) = \delta_{i,3} f_{\{1,2\}} 
\exp{(\theta_{\{1,2\}})}$, where $f_{\{1,2\}}$ are fixed amplitudes 
and $\theta_{\{1,2\}}$ are independent random phases, $\delta$-correlated in time. 
Random phases gives  an 
homogeneous statistics, without destroying the high anisotropy 
introduced by the chosen forced wavenumber.
We simulated the RKF at resolution $256^3$ and collected up 
to $70$ eddy  turn over times.

Anisotropy is studied by means of \SO\ decomposition of longitudinal structure functions\,:
\be
S_p\lp\BR\rp = \la\left[\lp \v\lp\x+\BR\rp-\v\lp\x\rp\rp \cdot \hat\BR\right]^p \ra,
\ee
where we have kept only the dependency on 
$\BR$ and neglected the small non-homogeneous fluctuations. 
We expect that the undecomposed structure functions are not the real 
``scaling'' bricks of the theory. 
Theoretical and numerical analysis showed \cite{alp00,abmp99,bt00} 
that one must first decompose the structure functions onto
irreducible representations of the rotational group and then study
the scaling behavior of the projections.
In practice, being the longitudinal structure functions scalar objects, 
their decomposition reduces to the projections on the spherical harmonics:
\begin{equation}
S_p(\BR) = \sum_{j=0}^{\infty}\sum_{m=-j}^{j} S_p^{jm}\lp\left|\BR\right|\rp Y_{jm}(\unitR).
\label{so3_sf}
\end{equation} 
As usual, we use indexes $(j,m)$ to label, respectively, the total angular momentum and its projection along a reference axis, say $\hat{\z}$.
The whole physical information is hidden in the functions $S_p^{jm}(R)$.
In particular, the main question we want to address here concerns their scaling properties:
\be
S_p^{jm}(|\BR|) \sim A_{jm} |\BR|^{\zeta^j(p)}.
\ee

First, we need an estimate for the ``dimensional'' values of the exponents 
$\zeta^j_d(p)$ in all sectors.
Our argument is based on the idea that large-scale energy pumping
and/or boundary conditions are such as to enforce a large-scale, 
anisotropic, driving velocity field $\U$. 
Dimensional predictions for intermediate (small) scales anisotropic 
fluctuations may then be obtained by studying the influence 
of the large-scale $\U$ on the inertial range. 
Let us therefore evaluate the weight of anisotropic contributions
as it comes out from a balance between inertial advection
of the small scales and the ``shear effect'', induced by the 
instantaneous large scale velocity configuration. 
Decomposing the velocity field in a small scale component, $\u$, and a large scale, strongly anisotropic component, $\U$, one finds the following equation 
for the time evolution of $\u$:
\begin{figure}[t]
\epsfxsize=\hsize
\hskip -.4cm
\epsfbox{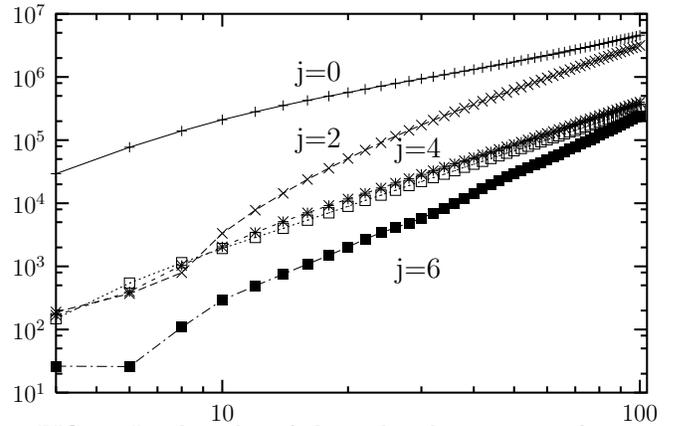}
\caption{Log-log plot of the 4-{\em th} order
 structure function projections, $|S_4^{jm}(R)|$, in all  sectors $(j,m)$ with a
high signal-to-noise ratio.
Notice that the isotropic sector remains the most energetic and  
the clear improving of scaling properties if measured on the 
projected quantities \protect\cite{bt00}.}
\label{fig:alls2}
\end{figure}
\be
\partial_t u_i + u_k\partial_k u_i + U_k\partial_k u_i + u_k\partial_k U_i = -\partial_i p + \nu\Delta u_i.
\label{eq:shear}
\ee
The major effect of the large-scale field is given by 
the instantaneous shear ${\cal S}_{ik} = \partial_k U_i$
which acts as an anisotropic forcing term
on small scales. 
A simple dimensional reasoning can be done as follows. 
Let us first consider the equation of motion
for two point quantities $\la u_l(\x')u_i(\x)\ra$ in the stationary 
regime; we may balance inertial terms and shear-induced terms
as follows:
\be
\la u_l(\x') u_k(\x)
 \partial_k u_i(\x) \ra \sim \la {\cal S}_{ik}(\x) u_l(\x') u_k(\x) \ra,
\label{eq:bal}
\ee 
which allows for a dimensional estimate of the anisotropic 
components of the LHS in terms of the RHS shear intensity and of the 
$\la uu \ra$ isotropic part. 
Similarly for three point quantities we have (neglecting tensorial details):
$\la uuu \partial u \ra \sim \la {\cal S} uuu \ra\label{eq:fourth}$
which can be easily generalized to any order velocity 
correlations. The shear term is a large-scale ``slow'' quantity and 
therefore, as far as scaling properties are concerned, we may safely estimate:
$\la {\cal S}_{ik}(\x) u_l(\x') u_k(\x) \ra 
\sim D_{ik} \la u_l(\x')u_k(\x) \ra$. Here the matrix $D_{ik}$ is associated 
to the combined probability to have a given shear and a given
small scale velocity configuration.
Clearly the $D_{ik}$ tensor brings angular momenta only up to $j=2$. 
One may therefore argue, by using simple composition of angular momenta, 
the following dimensional matching \cite{foot1}:
\be
\stackrel{j}{S^{j}_p(R)} \;\sim\;\; \stackrel{2\;\;\;\;\otimes\;\;\;\;j-2}{R\; {\cal S}\;\; \cdot \;\; S^{j-2}_{p-1}(R)}, 
\label{eq:gen}
\ee
where on top of each term we have written the total angular momentum of that 
contribution. $S^j_{p}(R)$ is the projection
on the $j$-{\it th} sector of the $p$-{\it th} order correlation function at scale $R$ (see equation \ref{so3_sf}), 
and ${\cal S}$ is the intensity of the shear term, $D_{ik}$, in the $j=2$ sector.
\begin{figure}[t]
\epsfxsize=\hsize
\hskip -.4cm
\epsfbox{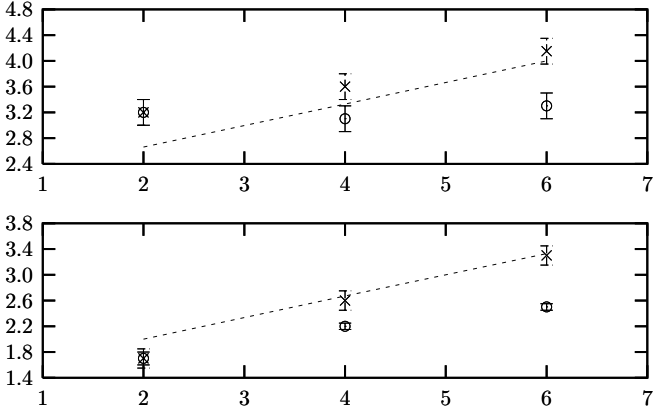}
\caption{Comparison between the 
dimensional estimate, $\zeta_d^j(p)=(p+j)/3$,
(straight lines), the measured
exponents, $\zeta^j(p)$ ($\circ$) and the exponents, $\zeta^j_r(p)$,
 obtained after random dephasing ($\times$), for
$p=2,4,6$. Top: sector $j=6$, bottom:
sector $j=4$.}
\label{fig:exp_sum}
\end{figure}
For instance, the leading behaviour of the 
$j=2$ anisotropic sector of the $3$-{\em th} order
 correlation function in the LHS 
of (\ref{eq:bal}) is given by the coupling between the $j=2$ components
of $D_{ik}$ and the $j=0$ sector of the  
$2$-{\em th} order velocity correlation in the RHS of (\ref{eq:bal}): $
S^2_3(R) \sim R\,{\cal S} \cdot S^0_2(R) \sim  R^{\zeta_d^{j=2}(3)}$. By using 
the same argument and considering that now
we know the scaling of $j=0$ and $j=2$ sectors of the third order correlation, 
we may estimate the scaling exponents of the fourth order correlation for $j=2,4$. 
From equation (\ref{eq:gen}), we have the dimensional matching in the $j=2$ sector: $
S_4^2(R)  \sim R\,{\cal S} \cdot S^0_3(R) \sim  R^{\zeta_d^{j=2}(4)} $ and
in the $j=4$ sector: $S_4^4(R)  \sim R\,{\cal S} \cdot S^2_3(R) \sim  R^2 {\cal S}^2 \cdot S^0_2(R) \sim R^{\zeta_d^{j=4}(4)}$.
The procedure is easily extended to higher orders:
\bea
\label{eq:dimpre2}\zeta_d^{j=2}(p) &=& \zeta_d^{j=0}(p-1) +1 = (p+2)/3, \;\;\;\; p>2;\\
\label{eq:dimpre4}\zeta_d^{j=4}(p) &=& \zeta_d^{j=2}(p-1) +1 = (p+4)/3,  \;\;\;\; p>3;\\
\label{eq:dimpre6}\zeta_d^{j=6}(p) &=& \zeta_d^{j=4}(p-1) +1 = (p+6)/3,  \;\;\;\; p>4;
\eea
which can be summarized as
$$ \zeta^j_d(p)={{(p+j)} \over 3},$$
 where intermittency effects in the isotropic sector have been neglected for simplicity. 
In this way, giving as input only the isotropic exponents,
 $\zeta_d^{j=0}(p)$,
we are able to predict the scaling exponents up to $j=2$ for the third order
structure functions, to $j=4$ for the fourth order, to $j=6$
for the fifth order and so on.
We may  do a little better by giving  a
prediction also for anisotropic fluctuations of second order correlation
functions. This cannot be simply obtained by using the equations
of motion, because the first one involving velocity correlations
at different spatial locations, i.e. inertial range quantities, 
is that for $\partial_t \la u_i(x) u_j(x') \ra$, 
which fixes a constraint only for the third order correlation function
(\ref{eq:bal}).
A way out is to ask the second order anisotropic fluctuations to be
analytic in the shear intensity, ${\cal S}$, consistently with what one finds 
for higher order structure functions by the above dimensional estimate. 
With this assumption, we recover for $j=2$ Lumley prediction \cite{lu67}, 
$\zeta^{j=2}_d(2)=4/3$ by simply writing the first two terms 
dimensionally consistent with an expansion in the shear intensity: $
\la uu \ra  \sim  (\epsilon R)^{2/3} + {\cal S} R^{4/3} + \dots$
where the first corresponds to the isotropic  scaling, while
the second captures anisotropies up to $j=2$ (higher $j$-sectors
could be captured by adding other terms in the expansions).
By using this  argument, we may now remove the limit of 
validity of the dimensional prediction, (\ref{eq:dimpre2}-\ref{eq:dimpre6}), 
and extend it to all $p$ values. \\
We now come to our numerical results for the \SO\ decomposition
of longitudinal structure functions.
In Figure \ref{fig:alls2}, we present for the $4$-{\it th} order
longitudinal structure function, an overview of all sectors $(j,m)$  
which have a signal-to-noise ratio high enough to ensure stable results.
Sectors with odd $j$s are absent due to the parity symmetry of our
observable. We measured anisotropic fluctuations up to
$j=6$. Scaling exponents can be measured in almost all sectors
except for $j=2$ where an annoying oscillation in the sign of
$S_4^{2,m}(|R|)$ prevents us from giving a quantitative statement.\\
We notice, as it is also summarized in Figure \ref{fig:exp_sum}, that
all measured exponents show a clear departure from the 
dimensional prediction. For example we measure in the
$j=4$ sectors the values: $\zeta^4(2)=1.65(5)$, $\zeta^4(4)=2.20(5)$, $\zeta^4(6)=2.55(10)$, and in the $j=6$ sector: $\zeta^6(2)=3.2(2)$, $\zeta^6(4)=3.1(2)$, $\zeta^6(6)=3.3(2)$.
This is a first clear sign that anisotropic 
scaling exponents are intermittent.\\ 
The {\it importance of being anomalous} 
does not stand on the exact values of the exponents, 
but on the connection between anomalous scaling and universality.
Indeed, if correlation functions in the inertial range
 are not given by a dimensional matching with the large-scale
shear, it means that they are fixed only by
the inertial part of the Navier-Stokes evolution. In other words, they should  
enjoy strong universality properties 
with respect to changes of the large-scale physics,
similarly to what happens to ``zero-modes'' responsible 
for anomalous scaling and universality in linear hydrodynamical problems 
\cite{rev}. \\
Such a statement can even be tested in a different way.

We have taken the stationary configurations of the RKF 
and randomly re-shuffled all velocity phases\,: $\hat{u}_i(\k) \rightarrow P_{il}(\k)\,
\hat{u}_l(\k)\,exp(i\,\theta_l(\k))$, where $P_{il}(\k)$ is the incompressibility projector and $\theta_l(\k)=-\theta_l(-\k)$.  
In this way we expect to filter out the dominant intermittent
fluctuations coming from the inertial evolution, or at least those
intermittent contributions connected to non-trivial phase organization.
The rationale of the above statement
comes from the observation that anomalous scaling,
in linearly advected hydrodynamical models, is connected 
to the existence of statistically preserved structures with highly 
complex geometrical properties \cite{cv}. 
\begin{figure}[t]
\epsfxsize=\hsize
\hskip -.4cm\epsfbox{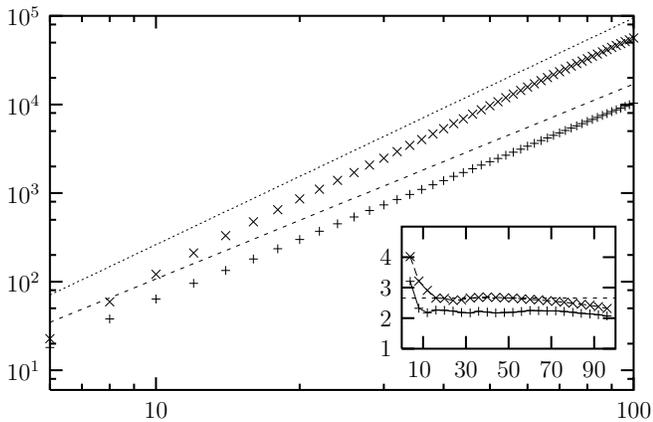}
\caption{Comparison of 
 scaling properties before ($+$) and after ($\times$) phases randomization
of the $4$-{\it th} order structure function
 for $j=4$. Straight lines are
the best fit slopes in the inertial range. 
In the inset, the changes for the logarithmic local 
slopes (same symbols); the horizontal dashed line corresponds to the 
dimensional prediction $\zeta_d^4(4)=8/3$.}
\label{fig:conpare}
\end{figure}
We imagine that once canceled the anomalous fluctuations,
the subdominant fluctuations, due to the dimensional balancing
with the forcing-shear terms, should show up.
Still, it is worth to remark, the statistics of the velocity field stays non-gaussian.\\
In Figure \ref{fig:conpare} we show the results for the decomposition 
of $4$-{\it th} order structure functions (after phase randomization) 
in the $j=4$ anisotropic sector. 
As it can be seen, scaling properties change 
significantly going from the anomalous value (before randomization) to 
the dimensional predictions (after randomization).\\
This happens for all sectors and moments we have measured, as it is summarized
in Figure \ref{fig:exp_sum}, with the notable exception of the second
order structure function where phase randomization has almost no effect.
An interesting fact which can have two explanations. Phases randomization is not enough to completely filter out  
intermittency, especially for two points quantities which should
be less sensible to phase correlation. Or, 
as noticed before, because second order correlation function is not 
constrained by any equation of motion, dimensional
 scaling may never exist for it even not as a sub-leading contribution.
This is an important point which certainly deserves further numerical, experimental tests.

In conclusions we have presented a dimensional argument able
to predict, by means of a matching between inertial correlations and 
shear-induced inertial terms, scaling exponents for all structure functions
in any anisotropic sector. 
We have shown by a direct numerical simulation that anisotropic scaling 
exponents deviate from the previous dimensional prediction, 
showing anomalous values.
When performing a random re-shuffling of all velocity phases, the dimensional scaling comes out as a sub-leading contribution.
Everything points toward the conclusion that anisotropic fluctuations
are anomalous and universal.

We thank I. Arad, M. Cencini and M. Vergassola for useful comments. 
This research was supported in part by the EU under the Grant 
No. HPRN-CT  2000-00162 ``Non Ideal Turbulence'' and by the INFM  
(Iniziativa di Calcolo Parallelo).

\end{multicols}

\begin{thebibliography}{99}
\bibitem{foot1} From now on, for the sake of simplicity, we neglect the 
dependency from the projection of the total angular momentum on the $z$ 
axis, associated to the  $m$ index.
\end{thebibliography}
\end{document}